\documentclass[twocolumn]{aastex61}

\shorttitle{Spectroscopic variations of V960 Mon}
\shortauthors{Takagi et al.}
\begin{document}

\title{The spectroscopic variations of the FU Orionis object V960 Mon}

\correspondingauthor{Yuhei Takagi}
\email{takagi@naoj.org}

\author{Yuhei Takagi}
\affil{Subaru Telescope, National Astronomical Observatory of Japan, 650 North A`ohoku Place, Hilo, HI 96720, USA}
\affil{Nishi-Harima Astronomical Observatory, Center for Astronomy, University of Hyogo, 407-2, Nishigaichi, Sayo, Sayo, Hyogo 679-5313, Japan}

\author{Satoshi Honda}
\affil{Nishi-Harima Astronomical Observatory, Center for Astronomy, University of Hyogo, 407-2, Nishigaichi, Sayo, Sayo, Hyogo 679-5313, Japan}

\author{Akira Arai}
\affil{Koyama Astronomical Observatory, Kyoto Sangyo University, Motoyama, Kamigamo, Kita-ku, Kyoto City, 603-8555, Japan}
\affil{Nishi-Harima Astronomical Observatory, Center for Astronomy, University of Hyogo, 407-2, Nishigaichi, Sayo, Sayo, Hyogo 679-5313, Japan}

\author{Kumiko Morihana}
\affil{Division of Particle and Astrophysical Science, Graduate School of Science, Nagoya University, Furo-cho, Chikusa-ku, Nagoya, Aichi, 464-8602, Japan}
\affil{Nishi-Harima Astronomical Observatory, Center for Astronomy, University of Hyogo, 407-2, Nishigaichi, Sayo, Sayo, Hyogo 679-5313, Japan}

\author{Jun Takahashi}
\affil{Nishi-Harima Astronomical Observatory, Center for Astronomy, University of Hyogo, 407-2, Nishigaichi, Sayo, Sayo, Hyogo 679-5313, Japan}

\author{Yumiko Oasa}
\affil{Faculty of Education, Saitama University, 255 Shimo-Okubo, Sakura, Saitama, Saitama 388-8570, Japan}

\author{Yoichi Itoh}
\affil{Nishi-Harima Astronomical Observatory, Center for Astronomy, University of Hyogo, 407-2, Nishigaichi, Sayo, Sayo, Hyogo 679-5313, Japan}



\begin{abstract}

We present the results of the spectroscopic monitoring of the FU Orionis type star V960 Mon. 
Spectroscopic variations of a FU Orionis type star will provide valuable information of its physical nature and the mechanism of the outburst. 
We conducted medium-resolution (R$\sim$10000) spectroscopic observations of V960 Mon with 2m Nayuta telescope at the Nishi-Harima Astronomical Observatory, from 2015 January to 2017 January for 53 nights in total. 
We focused on H$\alpha$ line and nearby atomic lines, and detected the strength variations in both absorption and emission lines. 
The observed variation in the equivalent width of the absorption lines correspond to a decrease in effective temperature and increase in surface gravity. 
These variations were likely to originate from the luminosity fading of the accretion disk due to the decrease in mass accretion rate.

\end{abstract}




\section{Introduction} \label{sec:intro}
Circumstellar materials accreting into a young star via protoplanetary disk form a photosphere. 
The luminosity of a pre-main sequence star shows an irregular variation due to the unstable mass accretion, clumpy materials in the protoplanetary disk, and presence of spots in the photosphere \citep[e.g.,][]{1994AJ....108.1906H}. 
In addition, these stars rarely undergo episodic outbursts, during which the visual brightness of the star increases by several (4--6) magnitudes. 
They are categorized as FU Orionis type stars (FUors).

The mechanism of outburst in the FUors is important for our understanding of the star formation process and the mass accretion. 
FUors outbursts are observed to occur during the Class I/II phase of the pre-main sequence stage \citep{2007ApJ...668..359Q}. 
The typical optical spectrum of FUors is similar to those of F- to G-type supergiants with broad absorption lines \citep{1977ApJ...217..693H}. 
By contrast, their spectrum in the near-infrared is similar to that of M-type supergiants \citep[e.g.,][]{1978ApJ...222L.123M,1992ApJ...398..273S}. 
Moreover, the profiles of their absorption lines are generally double-peaked \citep[e.g.,][]{1985ApJ...299..462H}. 
On the basis of these facts, the model that outbursts occur in the accretion disk, in which the mass accretion rate dramatically increases by several orders of magnitude, was proposed \citep{1996ARA&A..34..207H} and has been generally accepted. 
The spectroscopic characteristics of optical and near-infrared absorption lines in FUors are explained with the spectrum arising from the atmosphere of luminous accretion disk. 
The episodic outburst of the accretion rate could be the main factor to solve the ``luminosity problem" in low-mass protostars \citep{2014prpl.conf..387A}. 
On the other hand, there is a case that the spectroscopic characteristics are explained by a rapid rotating photosphere. \citet{2008AJ....136..676P} argued that the rapid rotating central star with large spots can explain the absorption profiles in optical wavelength, in the case of V1057 Cyg. 
Due to the rarity of FUors, well studied FUors are limited to several pre-main sequence stars (e.g., FU Ori, V1057 Cyg, V1515 Cyg). 
Therefore further studies of other FUors are necessary to understand the detail of their outburst mechanism.

Spectroscopic monitoring observations give valuable information to understand the physical nature of FUors. 
\citet{2012MNRAS.426.3315P} conducted a high-resolution spectroscopic monitoring of FU Ori and discussed the nature of the disk and the wind, investigating the periodic variation of the wing structure of the strong absorption lines. 
\citet{1998A&A...331L..53P} found the variations in the emission features of V1057 Cyg and discussed the formation and the evolution of the surrounding shell features. 
Since the spectroscopic and photometric characteristics of FUors vary among them, monitoring observations of other FUors are important. 

V960 Mon (2MASS J06593158-0405277) is an FU Ori type star identified in 2014 November \citep{2014ATel.6770....1M}. 
The progenitor is a typical Class II object with a spectral type of late K and a mass of 0.75$M_{\odot}$ \citep{2015ApJ...801L...5K}. 
Results of photometric observations were presented in several studies. 
According to the result of \citet{2015A&A...582L..12H}, the maximum magnitude of V960 Mon was observed on 2014 October 6, which was 10.8 and 10.2 mag in the Sloan $r$- and $i$-bands, respectively. 
The brightness of V960 Mon gradually decreased by 0.5 -- 1 mag in the optical wavelength between 2014 October and 2015 April. 
After V960 Mon had became observable at 2015 September, \citet{2015ATel.8019....1S} reported that the brightness of V960 Mon was comparable to that in April. 

Several spectroscopic observations were conducted after the outburst. 
\citet{2014ATel.6797....1H} conducted high-resolution spectroscopic observations and found that the optical spectrum matches that of early F-type giant or supergiant. 
In addition, the P Cygni profile in the H$\alpha$ line and the asymmetric profile of H$\beta$, Ca~{\footnotesize II} triplet, NaD doublet show evidence of outflow. 
Li absorption are also seen in its spectrum. 
From the near-infrared spectroscopic observations \citep{2015ATel.6901....1P,2015ATel.6862....1R}, CO absorption bands and broad ${\mathrm H}_2{\mathrm O}$ absorptions were found, which are characteristic of late-type stars. 
The spectral type inconsistency between optical and near-infrared is a well-known characteristic of FUors \citep{1996ARA&A..34..207H,2016NewA...43...87J}. 
Therefore V960 Mon is classified as FUors in previous studies, however it is important to keep in mind that it may be a different type of an eruptive star compared to a typical FUors due to its small brightness increase.

We conducted medium-resolution optical spectroscopic monitoring observations of V960 Mon with the Nayuta Telescope in Nishi-Harima Astronomical Observatory. 
The spectra of 6300 -- 6750~{\AA} were obtained from 2015 January to 2017 January. 
The details of the observations are described in section 2. 
The variations found in both absorption lines and emission lines are summarized in section 3. 
In section 4, we discuss the mechanism of the spectroscopic variations.

\renewcommand{\arraystretch}{0.85}
\begin{table*}[t]
\caption{Observation Log of V960 Mon.} \label{tab:log}
\begin{center}
\begin{tabular}{lcrr}
   \hline \hline
   \multicolumn{1}{c}{Observation date}	& Average MJD	& \multicolumn{1}{c}{Exp. time}	& \multicolumn{1}{c}{S/N}	\\ 
   \multicolumn{1}{c}{(JST)}		& 		& \multicolumn{1}{c}{(sec)}	& 	\\ \hline
   2015 Jan 27  & 57049.5947    & 1800  & 43.2 \\
   2015 Jan 28  & 57050.5101    & 3900  & 57.4 \\
   2015 Feb 14  & 57067.6564    & 3000  & 60.7 \\
   2015 Feb 15  & 57068.5432    & 4800  & 35.5 \\
   2015 Feb 18  & 57071.4059    & 600   & 19.5 \\
   2015 Mar 02  & 57083.5397    & 3300  & 79.6 \\
   2015 Mar 08  & 57089.4436    & 2100  & 38.3 \\
   2015 Mar 11  & 57092.5300    & 3000  & 47.5 \\
   2015 Mar 12  & 57093.4212    & 2700  & 59.6 \\
   2015 Mar 14  & 57095.4201    & 3000  & 54.4 \\
   2015 Mar 22  & 57103.4784    & 3000  & 41.3 \\
   2015 Mar 24  & 57105.5763    & 3000  & 23.5 \\
   2015 Mar 25  & 57106.5372    & 2400  & 28.0 \\
   2015 Mar 26  & 57107.5559    & 3600  & 54.3 \\
   2015 Mar 27  & 57108.4223    & 1800  & 34.7 \\
   2015 Mar 28  & 57109.4830    & 4200  & 58.1 \\
   2015 Mar 29  & 57110.4883    & 4800  & 45.0 \\
   2015 Mar 30  & 57111.4818    & 4200  & 55.2 \\
   2015 Apr 08  & 57120.4800    & 4200  & 28.5 \\
   2015 Apr 27  & 57139.4747    & 3600  & 55.4 \\
   2015 Sep 14  & 57279.7965    & 5400  & 60.8 \\
   2015 Oct 03  & 57298.8257    & 3600  & 37.7 \\
   2015 Oct 06  & 57301.7157    & 1800  & 20.1 \\
   2015 Oct 07  & 57302.8234    & 3000  & 22.1 \\
   2015 Oct 09  & 57304.7782    & 1800  & 45.8 \\
   2015 Oct 13  & 57308.8442    & 3000  & 59.5 \\
   2015 Oct 14  & 57309.7998    & 3600  & 49.8 \\
   2015 Oct 16  & 57311.8349    & 3000  & 49.1 \\
  \hline
  \end{tabular}
\begin{tabular}{lcrr}
   \hline \hline
   \multicolumn{1}{c}{Observation date}	& Average MJD	& \multicolumn{1}{c}{Exp. time}	& \multicolumn{1}{c}{S/N}	\\ 
   \multicolumn{1}{c}{(JST)}		& 		& \multicolumn{1}{c}{(sec)}	& 	\\ \hline
   2015 Oct 23  & 57318.8292    & 3000  & 66.5 \\
   2015 Oct 30  & 57325.7795    & 13200 & 61.1 \\
   2015 Oct 31  & 57326.7295    & 12600 & 70.1 \\
   2015 Nov 02  & 57328.8333    & 3600  & 37.0 \\
   2015 Nov 04  & 57330.8372    & 3600  & 81.6 \\
   2015 Nov 10  & 57336.8470    & 3000  & 48.6 \\
   2015 Nov 20  & 57346.8470    & 3000  & 43.6 \\
   2015 Nov 26  & 57352.7979    & 5400  & 51.1 \\
   2015 Nov 29  & 57355.7595    & 4800  & 95.4 \\
   2015 Nov 30  & 57356.8376    & 7200  & 80.1 \\
   2015 Dec 22  & 57378.5756    & 6000  & 54.2 \\
   2015 Dec 27  & 57383.6180    & 6000  & 56.1 \\
   2016 Jan 04  & 57391.6662    & 8400  & 56.6 \\
   2016 Jan 06  & 57393.6529    & 6000  & 58.7 \\
   2016 Jan 08  & 57395.6807    & 12000 & 81.5 \\
   2016 Jan 10  & 57397.7164    & 6000  & 24.2 \\
   2016 Jan 15  & 57402.5282    & 6000  & 62.3\\
   2016 Mar 01  & 57448.4884    & 1200  & 40.1 \\
   2016 Mar 02  & 57449.5273    & 7200  & 106.6 \\
   2016 Mar 28  & 57475.5007    & 9600  & 104.5 \\
   2016 Apr 08  & 57486.5306    & 3000  & 28.3 \\
   2016 Apr 15  & 57493.5308    & 3600  & 27.5 \\
   2016 Apr 19  & 57497.5273    & 3000  & 21.6 \\
   2016 Nov 25  & 57717.8062    & 2400  & 35.2 \\
   2016 Dec 11  & 57733.7138    & 3600  & 31.2 \\
   2017 Jan 10  & 57763.5014    & 6000  & 47.3 \\
   2017 Jan 31  & 57784.4518    & 8400  & 53.3 \\
   &&&\\
   \hline
  \end{tabular}
 \end{center}
\end{table*}
\renewcommand{\arraystretch}{1}

\section{Observations and data reductions} \label{sec:obs}
The spectroscopic observations of V960 Mon were conducted with the Medium And Low-resolution Longslit Spectrograph \citep[MALLS;][]{Ozaki} equipped on the Nasmyth focus of the 2m Nayuta telescope at the Nishi-Harima Astronomical Observatory. 
Three gratings (150, 300, and 1800~l/mm) and five slits with different widths (0.8, 1.2, 1.6, 5.0, and 8.0 arcsec) are mounted on this spectrograph. 
We used the 1800~l/mm grating and the 0.8~arcsec slit to achieve the resolution power of R$\sim$10000 at 6500~{\AA}. 
The dispersed light was corrected with a 2K $\times$ 2K back-illuminated CCD detector (15 $\mu$m pixel) with a spectral coverage of $\sim$ 450~{\AA}. 
The grating angle was set to a proper amount to obtain the spectral features of the H$\alpha$ line (6562.8~{\AA}), Li absorption (6707.8~{\AA}), and several neighboring atomic lines, such as Fe~{\footnotesize I}, Fe~{\footnotesize II}, and Ca~{\footnotesize I}. 
The total observation nights were 53, between 2015 January 27 and 2017 January 31. 
The exposure time in a single frame was from 300 sec to 1200 sec. 
Exposures were repeated several times in order to improve the signal-to-noise ratio (S/N). 
Since an image rotator is not installed on the Nasmyth focus and the MALLS, the position angle of the slit was different in each exposure.
The typical seeing of the observations were 1.5 arcsec, and therefore the spectrum of V960 Mon were obtained by separating the close-by star 2MASS J06593168-0405224 with a separation of 5.7 arcsec \citep{2015ApJ...801L...5K}.
Table 1 shows the observation log. 
For comparison, the spectrum of $\beta$~Aqr (G0Ib, $\sim$5500K) was also obtained as a standard star on 2015 June 16 with the same instrumental settings, because the smoothed $\beta$~Aqr spectrum is known to be comparable to those of FUors.

\begin{figure*}[t!]
 \begin{center}
   \includegraphics[width=16cm]{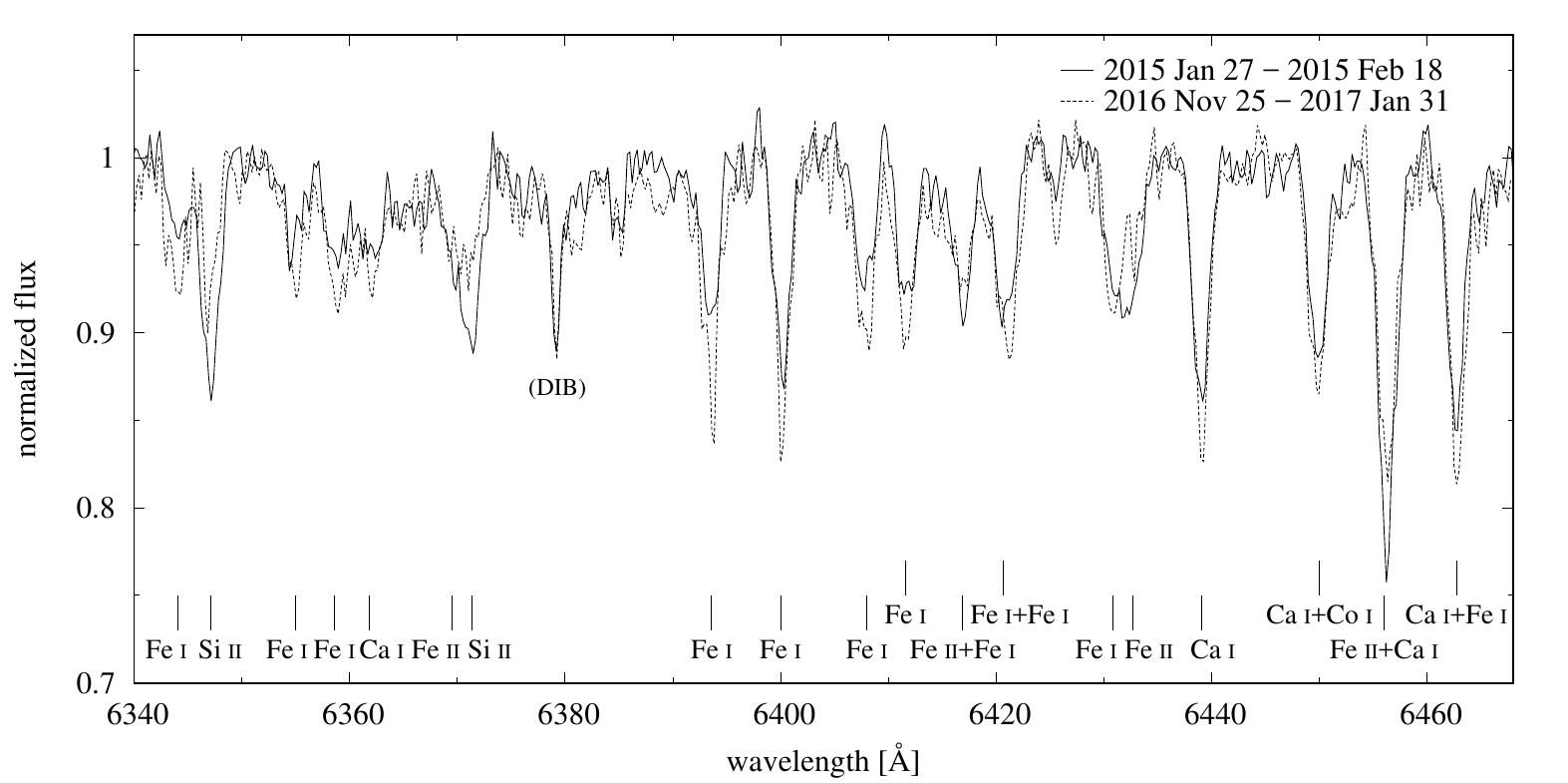} 
 \end{center}
 \caption{Comparison of the V960 Mon spectrum taken in different phases. The black solid line shows the combined spectrum of 2015 January 27, 28, February 14, 15, and 18 (early phase). The dashed line is the combined spectrum made with the data of 2016 November 25, December 11, 2017 January 10, and 31 (late phase).}
 \label{fig:speccomp}
\end{figure*}

Data reductions were performed with the Image Reduction and Analysis Facility (IRAF) software package\footnote{IRAF is distributed by the National Optical Astronomy Observatory.}. 
Overscan subtraction, dark subtraction, cosmic ray rejection, flat fielding, distortion correction, and scattered light subtraction were conducted before the spectrum was extracted. 
After the wavelength were corrected to heliocentric wavelength, we estimated the radial velocity of V960 Mon.
We compared the observed peak wavelength of Fe~{\footnotesize I} 6393.6, Fe~{\footnotesize I} 6411.6, and Ca~{\footnotesize I} 6439.1 lines to those rest wavelength to calculate the radial velocity. The estimated radial velocity was 38.1$\pm$0.5 km ${\mathrm s}^{-1}$.
Continuum normalization was conducted after all the spectra collected in a single night were combined into a single spectrum.
The S/N was calculated from the continuum regions, which we set at the wavelength bands of 6616 -- 6622~{\AA}, 6650 -- 6659~{\AA}, and 6683 -- 6694~{\AA}. 
The S/Ns of the extracted spectra are shown in Table \ref{tab:log}. 

\section{Results} \label{sec:res}

During our observation campaign spanning two years, V960 Mon showed a long-term photometric and spectroscopic variations. 
According to the photometric data of the monitoring observations produced by the American Association of Variable Star Observers (AAVSO), the brightness of V960 Mon decreased by 1.07$\pm$0.01 and 0.79$\pm$0.01 mag in the $V$- and $I$-bands, respectively, from 2015 January 27 to 2017 January 31. 
In the following subsections, we focus on the variation of intensity of some absorption and emission lines of V960 Mon. 
We combined the spectra taken for several days to several weeks into a single spectrum to achieve high S/Ns ($>$50). 
The total number of the created combined spectra were 21. All the investigations and measurements mentioned in the following sections are based on these combined spectra.

\subsection{Absorption lines} \label{sec:res_ab}

The typical FU Ori type star has broad atomic absorption lines in its spectrum, and the strengths of these lines are similar to those in the spectra of F-G type supergiants. 
The averaged spectrum of V960 Mon over the full two years observations were comparable to those of a spun-up $\beta$ Aqr spectrum with a rotational velocity of $\sim$50 km ${\mathrm s}^{-1}$. 
On the other hand, the comparison of combined spectra of ``early phase" (2015 January 27, 28, February 14, 15, and 18) and the ``late phase" (2016 November 25, December 11, 2017 January 10 and 31) showed that the peak depths of absorption lines varied during our observation period (Figure \ref{fig:speccomp}). 
The peak depth of neutral atomic lines such as Fe~{\footnotesize I} and Ca~{\footnotesize I} in the late phase were deeper than those in the early phase spectrum. The peak depth increased by up to 10 \%.
Meanwhile, the peak depth of the ionized Fe lines in the late phase spectrum were shallow compared to those in the early phase spectrum. 
Their peak depth dropped by $\sim$5\% in the late phase. 
Since the peak depth of the neutral and ionized lines showed the opposite trends in variation, their variations cannot be explained only with the continuum excess fluctuation, such as veiling, which is frequently observed in pre-main sequence stars. 

\begin{figure*}[t!]
\plotone{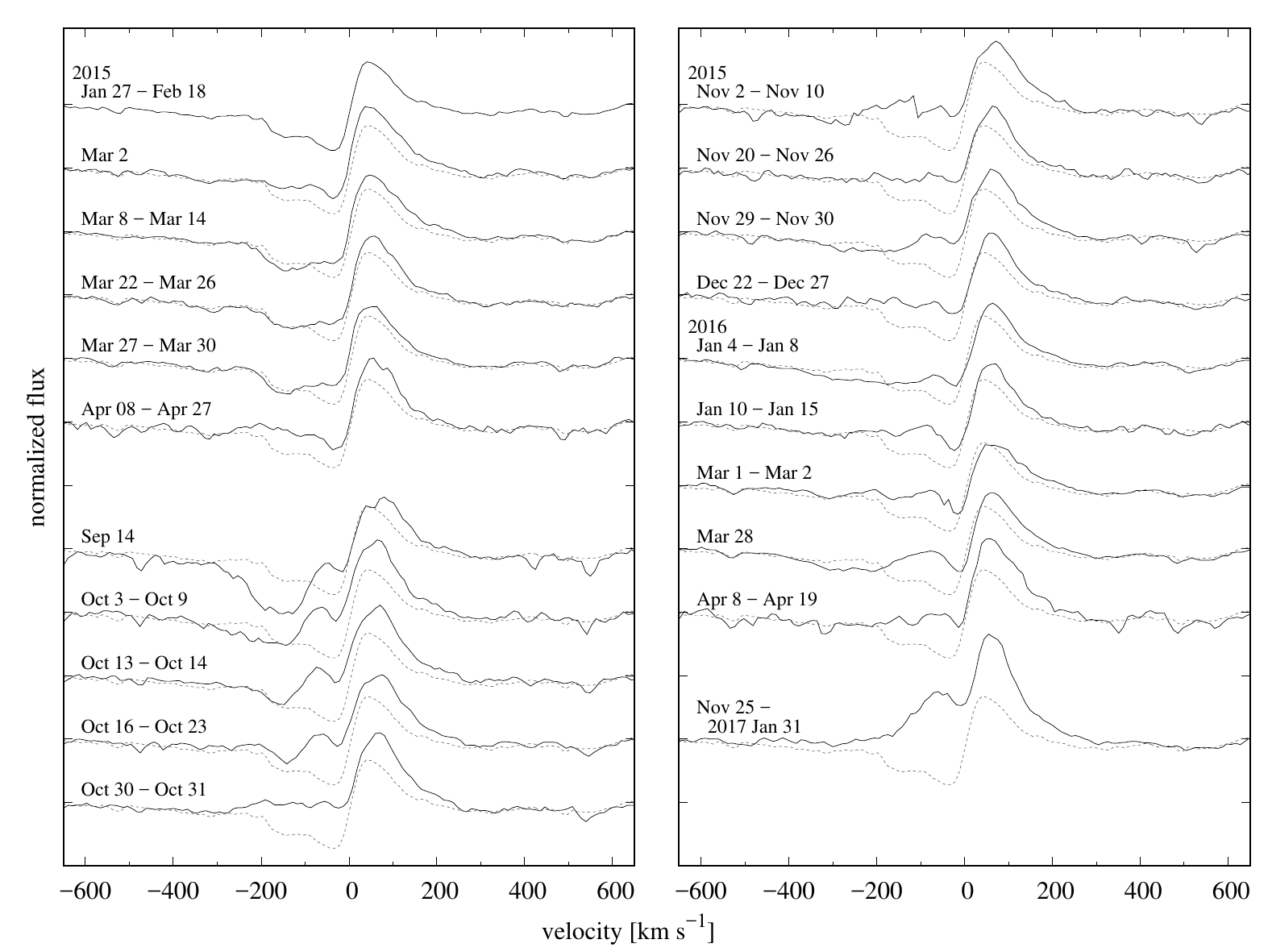}
\caption{Variations of the H$\alpha$ line. The dotted lines show the average spectra from 2015 Jan 27 to Feb 18 for comparison. \label{fig:Ha}}
\end{figure*}

\subsection{H$\alpha$ line} \label{sec:res_ha}

The P Cygni profile in the H$\alpha$ line appears in most of the obtained spectra (Figure \ref{fig:Ha}). 
We interpret that the H$\alpha$ line constitutes several components, such as blueshifted absorption feature due to the outflow, absorption near the velocity~(\textit{v}) of 0~km ${\mathrm s}^{-1}$, which corresponds to the atmospheric absorption similar to other atomic absorption lines, and the red emission. 
The time-variation in the blueshifted absorption indicates the evolution of the intense wind. 
\citet{2014ATel.6797....1H} reported  a variation in blueshifted absorption in strong lines (e.g., H$\alpha$, H$\beta$, Na D lines) with high-resolution spectroscopic observations for two nights (2014 December 9 -- 10). 
A large fluctuation in the blueshifted absorption of H$\alpha$ was also found in our spectra, which was especially prominent in the spectrum obtained from 2015 January to 2015 October. 
The blue edge of this absorption extended to $-$200 -- $-$300 km ${\mathrm s}^{-1}$, and the main peak was at roughly $-$150 km ${\mathrm s}^{-1}$. 
In the latter phase of the observation, the blueshifted absorption was difficult to resolve. 
Meanwhile, the intensity of the redshifted emission gradually increased during the observation period with a nearly constant peak velocity.

\begin{figure}[t!]
 \begin{center}
   \includegraphics[width=8cm]{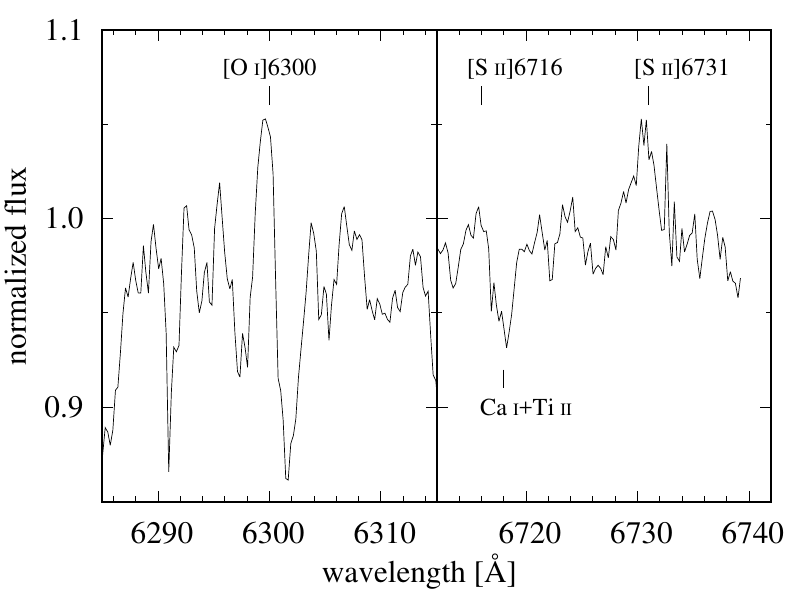} 
 \end{center}
 \caption{Forbidden lines of V960 Mon observed on 2016 March 2, of which the spectrum had the highest S/N taken in a single day in our observations.}
\label{fig:em}
\end{figure}

\begin{figure*}[t!]
\plotone{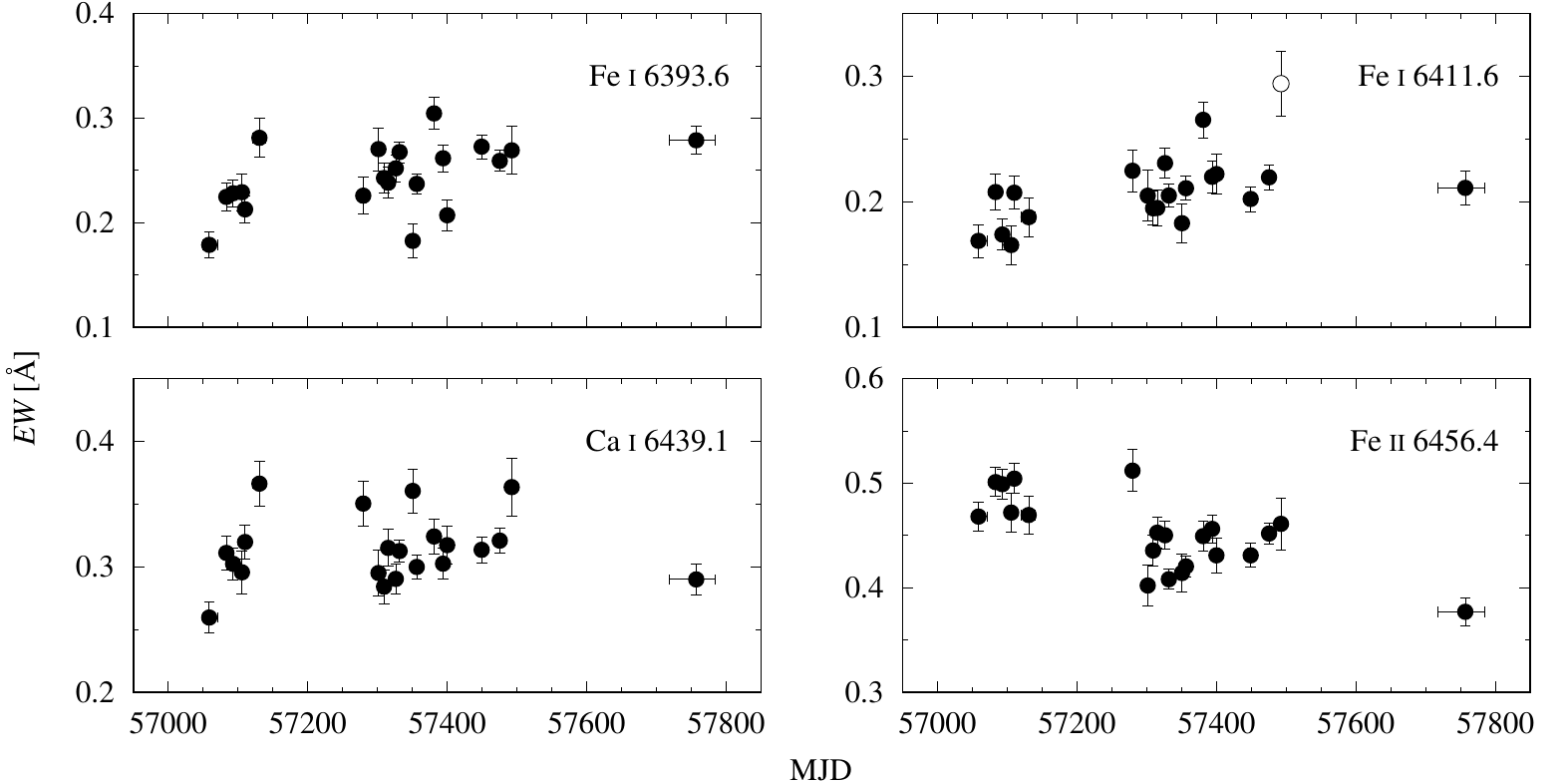}
\caption{\textit{EW}s of the absorption lines. The error bar in the x-axis shows the periods when the spectra were collected to create a combined spectrum for \textit{EW} measurement. The open circle indicates the \textit{EW} measured with a poor fitting due to a low-S/N (S/N $\sim$ 45). \label{fig:ew} }
\end{figure*}

\subsection{Forbidden lines} \label{sec:res_fb}

Some of the weak forbidden lines were also detected in our spectra (Figure \ref{fig:em}). 
The [S~{\footnotesize II}]6731 line was clearly observed. 
The equivalent widths (\textit{EW}s) of the [S~{\footnotesize II}]6731 was measured in all the combined spectra, and the derived averaged value was -0.093 $\pm$ 0.035~{\AA}. 
No significant strength variation was found. 
By contrast, [O~{\footnotesize I}]6300 line was hardly detected, presumably because it is intrinsically weak and is blended with the telluric lines. 
The [S~{\footnotesize II}]6716 line was not detected due to the nearby absorption of Ca~{\footnotesize I} and Ti~{\footnotesize II} in 6718~{\AA}. 
However, the fact that the observed line profiles of the blended Ca~{\footnotesize I} and Ti~{\footnotesize II} line were not symmetric suggests the presence of the neighboring [S~II]6716 emission. 
These forbidden lines have been rarely observed in FUors \citep[e.g., V2494 Cyg;][]{2013MNRAS.432.2685M}. 

\section{Discussion} \label{sec:dis}

\subsection{Variations of the absorption lines} \label{sec:dis_var}

To investigate the strength time-variations of the absorption lines during the two year observation period, we measured the \textit{EW}s of each of the 21 combined spectra. We used four lines which were not heavily blended with nearby lines: 
Fe~{\footnotesize I} 6393.6, Fe~{\footnotesize I} 6411.6, Ca~{\footnotesize I} 6439.1, and Fe~{\footnotesize II} 6456.4.
Nevertheless, some weak absorption lines exist close to the three atomic lines. 
Therefore, note that the measured \textit{EW}s of these lines are the sum of the target atomic line and the adjacent lines. 
Moreover, Fe~{\footnotesize II} 6456.4 was blended with nearby Ca {\footnotesize I}. 
However, we adopted this line for the \textit{EW} measurement because the trends in variation were contrastive between the neutral and ionized lines.
We fitted each line profile with a Gaussian function, using the SPLOT task of IRAF, and measured the \textit{EW}. 
The error of the equivalent width ($\delta$\textit{EW}) was determined using the equation given by \citet{1988IAUS..132.....C}, $\delta EW = 1.6(w\delta x)^{1/2}\epsilon$, where $w$, $\delta x$, and $\epsilon$ are the width of the line, pixel size, and reciprocal of S/N, respectively. 

We found that the \textit{EW} of Fe~{\footnotesize II} 6456.4 gradually decreased during the two-year monitoring (Figure \ref{fig:ew}). 
The \textit{EW}s of the Fe~{\footnotesize I} lines were nearly stable, or increased marginally. 
The dispersion in each \textit{EW} may imply a short-term variation, however a high time-resolution observations are needed to confirm this variation. 
The FWHM of each combined spectrum were estimated by averaging the measured FWHM of Fe~{\footnotesize I} 6393.6, Fe~{\footnotesize I} 6411.6, and Ca~{\footnotesize I} 6439.1 lines.
The FWHM was constant during the observation period, and the averaged FWHM of the entire period was 1.9$\pm$0.1~{\AA}.
We also measured the half-width at half-depth \citep[HWHD,][]{2008AJ....136..676P,2015ApJ...807...84L} of both the shortward and longward side of Fe~{\footnotesize I} 6393.6, Fe~{\footnotesize I} 6411.6, Ca~{\footnotesize I} 6439.1, and Fe~{\footnotesize II} 6456.4 lines in each combined spectrum, to figure out the presence of the variations in line profile. 
The HWHD of shortward and longward side were nearly equal each line during the observation period.
Therefore, wind-driven blueshifted absorption features which were detected in several lines of HBC 722 \citep{2015ApJ...807...84L} were not seen in these four lines. 
Due to their symmetric profiles, we interpret that these lines arise from the rapid rotating atmosphere and the spectroscopic variations originate from the change in effective temperature ($T_{\mathrm{eff}}$) and surface gravity ($g$) in the atmosphere.

\subsection{Estimating the $T_{\mathrm{eff}}$ and $g$ variations} \label{sec:dis_tg}

A general method to figure out the value of $T_{\mathrm{eff}}$, $g$, and also the metal abundance from a spectrum is based on \textit{EW}s measurements of Fe~{\footnotesize I} and Fe~{\footnotesize II} lines, taking into account the equality of the abundance estimated from Fe~{\footnotesize I} and Fe~{\footnotesize II}. 
Several tens of absorption lines are usually used for this investigation, to estimate $T_{\mathrm{eff}}$ and $g$ with no dependency on depths and excitation potentials of absorption lines. 
However, since the wavelength range ($\sim$~450~{\AA}) of our spectra is limited, it is difficult to make an accurate estimate of these parameters with the standard method. 
In order to estimate the approximate $T_{\mathrm{eff}}$ and $g$ in the spectra of early phase and the late phase (as defined in section \ref{sec:res_ab}), we first derived the \textit{EW} variability of the Fe~{\footnotesize I} 6393.6, Fe~{\footnotesize I} 6411.6, Ca {\footnotesize I} 6439.1, and Fe~{\footnotesize II} 6456.4 lines using the synthetic spectrum generated with an atmospheric model. 

We used the software SPTOOL\footnote{http://optik2.mtk.nao.ac.jp/\~{}takeda/sptool/} developed by Y. Takeda to create a synthetic spectrum, which is based on Kurucz's ATLAS9/WIDTH9 atmospheric model \citep{1993KurCD..13.....K}.  
We first assumed that the continuum excess did not change between the early and the late phases.
The metal abundance was set to the solar value, and the oscillator strength (\textit{gf}) of the four lines were calibrated to fit the solar spectrum presented by \citet{2000vnia.book.....H}. 
Synthetic spectra were created with SPTOOL by varying the parameters $T_{\mathrm{eff}}$ from 4500 K to 7500 K and log $g$ from 1.0 to 3.6. 
Microturbulence ($\xi$) were set to a value calculated by the following empirical relations among $T_{\mathrm{eff}}$, $g$, and $\xi$ from the result of \citet{2001AJ....121.2159G}:
\begin{eqnarray}
  \; \xi = &6.02g^{-0.08} \; &(\mathrm{for}\; T_{\mathrm{eff}}=7000 - 7500 K) \nonumber \\
  \; \xi = &10.58g^{-0.17} \; &(\mathrm{for}\; T_{\mathrm{eff}}=6500 - 7000 K) \nonumber \\
  \; \xi = &12.57g^{-0.23} \; &(\mathrm{for}\; T_{\mathrm{eff}}<6500 K), \nonumber
\end{eqnarray}
which were derived from the observed physical parameters of late A- to early G-type stars. 

\begin{figure*}[t!]
 \begin{center}
   \includegraphics[width=16cm]{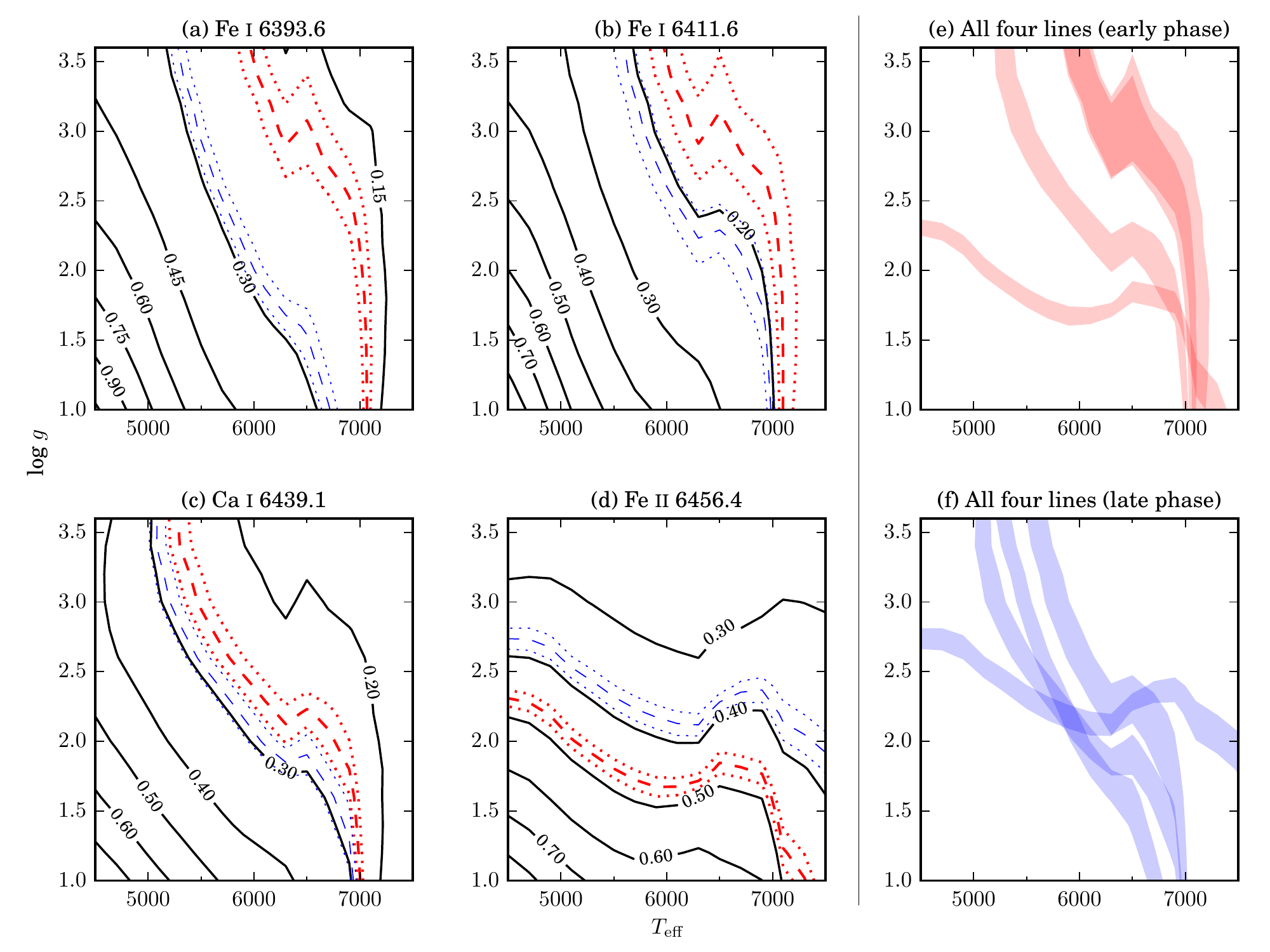} 
 \end{center}
 \caption{\textit{(a)--(d)}: Estimated \textit{EW}s of the four absorption lines from the synthetic spectra (black solid line). Thick red and thin blue dashed lines represent the \textit{EW}s of the combined spectra of V960 Mon obtained in early and late phases of the monitoring, respectively. The dotted lines indicate the errors of the \textit{EW}s. \textit{(e), (f)}: All the four red and blue lines in the left and middle panels are superposed in the respective frames ((e): early phase, (f): late phase). The filled region of each line shows to the error range. 
}
 \label{fig:cont}
\end{figure*}

The \textit{EW} of Fe~{\footnotesize I} 6393.6, Fe~{\footnotesize I} 6411.6, Ca {\footnotesize I} 6439.1, and Fe~{\footnotesize II} 6456.4 lines in the created synthetic spectrum were measured with the SPLOT task of IRAF. 
As mentioned in section \ref{sec:dis_var}, the absorption lines of V960 Mon are broad and the measured \textit{EW}s of four lines also include the components of adjacent lines. 
Therefore, prior to the \textit{EW} measurements of synthetic spectra, we first defined the wavelength range for each four absorption lines.
This wavelength range were determined based on the blue-end and the red-end of the absorption line in the observed V960 Mon spectrum. 
The \textit{EW} measurement of the four lines in the synthetic spectrum were conducted by summing up the absorption components included in this wavelength range, and then the variabilities of these lines were estimated (Figure \ref{fig:cont}).
Note that the bumps that appear in the temperature range of $T_{\mathrm{eff}}\sim$~6500~K are due to the switch of the equations adopted for $\xi$ estimations. 
The $T_{\mathrm{eff}}$ and log~$g$ can be estimated from the convergence of the four lines shown in the panel (e) and (f) of Figure \ref{fig:cont}.
The approximate parameters for V960 Mon in the early phase were estimated to be $T_{\mathrm{eff}}\sim$~7000~K and log~$g\sim$~1.2--1.5. 
In the late phase, $T_{\mathrm{eff}}$ decreased to $\sim$~5800 -- 6000~K and log~$g$ increased to $\sim$~2.2. 
Although the estimated $T_{\mathrm{eff}}$ and $g$ have uncertainties due to the limited number of absorption lines and the $\xi$ estimated from the empirical relationship, a trend of decrease in $T_{\mathrm{eff}}$ and increase in $g$ was shown from our observed spectra.

\textit{EW} variations may occur by the change in continuum excess such as veiling, therefore there is a possibility that the \textit{EW} variations may not be only caused by $T_{\mathrm{eff}}$ and $g$ variations.
Since the \textit{EW} of Fe~{\footnotesize I} show an increasing trend and Fe~{\footnotesize II} decreased during our observation period, it is difficult to explain these \textit{EW} variations with the fluctuation in the veiling, but we conducted an alternative investigation to derive the $T_{\mathrm{eff}}$ and $g$ variations. 
An unequivocal way to resolve the variability of the stellar parameters from \textit{EW} variations with no contamination of veiling is to use the \textit{EW} ratios among the lines. 
\textit{EW} ratios calculated with the lines within a limited wavelength are fine indicators of the physical parameters since they are free from veiling \citep{2014PASJ...66...88T,2015PASJ...67...87T}. 
\textit{EW} ratios composed of atomic line and ionized line (Fe~{\footnotesize I} 6393.6/Fe~{\footnotesize II} 6456.4, Fe~{\footnotesize I} 6411.6/Fe~{\footnotesize II} 6456.4, and Ca {\footnotesize I} 6439.1/Fe~{\footnotesize II} 6456.4) were employed as indicators. 
By using the measured \textit{EW}s of Fe~{\footnotesize I}, Fe~{\footnotesize II}, and Ca~{\footnotesize I} lines of the synthetic spectra, we estimated the relationship between \textit{EW} ratio, $T_{\mathrm{eff}}$, and $g$, as the case of \textit{EW}. 
Then the variations of $T_{\mathrm{eff}}$ and $g$ of V960 Mon were estimated by comparing its \textit{EW} ratio of the early and late phase spectrum to the derived \textit{EW} ratio - $T_{\mathrm{eff}}$ - $g$ relations. 
All of the \textit{EW} ratios show the decrease in $T_{\mathrm{eff}}$, which was comparable to the result estimated from the \textit{EW} variations. 
On the other hand, it was difficult to confirm the increase in $g$.
Therefore an investigation using more absorption lines are necessary to derive the precise variations of these parameters.

We constrained the cause of the spectroscopic and photometric variations based on $T_{\mathrm{eff}}$ and $g$ variations estimated from the \textit{EW} variation. 
According to the previous photometric observations \citep{2015A&A...582L..12H}, $T_{\mathrm{eff}}$ decrease and $g$ increase occurred with a decrease in brightness. 
These variations are difficult to be explained with a photospheric pulsation of a variable star, or a gravitational contract of a photosphere of young stars. 
The pulsation of star photosphere will cause the variations of brightness, $T_{\mathrm{eff}}$, and $g$.
However, the brightness of the star decreases when the radius of the photosphere increases, resulting in a decrease in both $T_{\mathrm{eff}}$ and $g$. 
This variation conflicts with that of V960 Mon case. 
Meanwhile, photospheric contraction in pre-main sequence stage cause $T_{\mathrm{eff}}$ and $g$ variation. However, it is again difficult to explain the $T_{\mathrm{eff}}$ and $g$ variations of V960 Mon with this photospheric contraction, since this variation underwent in such a short timescale. 

We interpret that the absorption lines in the optical spectrum of V960 Mon arise from the atmosphere of the accretion disk. 
The decrease in $T_{\mathrm{eff}}$ during our observation period is the evidence of the decrease in the accretion rate. 
The decreasing trends in both the brightness and $T_{\mathrm{eff}}$ are similar to the variation found in V1057 Cyg. 
According to the steady accretion disk model, where the disk is considered to be a luminous source, the decrease in brightness and $T_{\mathrm{eff}}$ can be explained by the decreasing mass accretion rate \citep{1988ApJ...325..231K,1991ApJ...383..664K}. 
During our observation period, the $T_{\mathrm{eff}}$ derived from our optical spectra decreased from $\sim$7000~K to $\sim$6000~K.
The brightness of V960 Mon also decreased during the observation period but still bright compared to the pre-outburst phase.
Therefore, the decrease in both $T_{\mathrm{eff}}$ and brightness is considered to be caused by the decrease of the accretion rate. 
The mass accretion rate of V960 Mon reduced nearly 50\% from the early phase to the late phase, based on the accretion disk model and the observed $T_{\mathrm{eff}}$ variation. 
In addition, the increase in $g$ suggests the evolution of the vertical structure in the accretion disk. 
\citet{1990MNRAS.242..439C} showed that the scale-height of the accretion disk gradually decreases with the disk cooling. 
This trend is comparable to the variations in both $T_{\mathrm{eff}}$ and $g$ observed in V960 Mon. 
We conclude that the outburst and the brightness decrease that we observed are the result of a sudden increase and decrease in the mass accretion rate, respectively, in V960 Mon. 

\subsection{Variation in H$\alpha$ line} \label{sec:dis_ha}

The H$\alpha$ line show a P Cygni profile especially in the spectra obtained from 2015 January to 2015 October. 
The blueshifted absorption feature became weak in the latter phase of the observation period. 
The decline of the blueshifted absorption feature is comparable to those observed in H$\beta$ and NaD lines in HBC 722 \citep{2015ApJ...807...84L}. 
On the other hand, strong emissions are rare in FUors. 
Indeed, the emission part of the H$\alpha$ line was hardly detected in one of the earliest spectra \citep{2014ATel.6770....1M}. 
The increasing trend implied in the peak intensity of H$\alpha$ emission (Figure \ref{fig:Ha}) may be the result of scale-height evolution of the V960 Mon disk, since it is considered that the increase in the scale-height during the FU Ori outbursts is one of the solution of the lack of the ultraviolet fluxes and emission lines from the boundary layer \citep{1996ARA&A..34..207H}.  
However, because H$\alpha$ line is composed of several components (section \ref{sec:res_ha}), it is difficult to investigate the accurate strength variation of the H$\alpha$ emission. 
Therefore, we removed the absorption component near $v=0$ km ${\mathrm s}^{-1}$ which is considered to originate from the disk atmosphere, using the spectrum of field giants. 
By approximating the disk atmosphere spectrum by the field giant spectrum, and subtracting it from the observed V960 Mon spectrum, it is able to remove the spectroscopic feature of the disk atmosphere from the observed spectrum. 
Although the spectrum of V960 Mon includes both the photospheric and disk components, the latter component is dominant in the spectrum. 
Therefore we assumed that giant star spectrum fitting well the observed V960 Mon spectrum as the ``disk component" of V960 Mon.

We searched the archival data for field giant stars of which the atomic absorption lines in its spectra fit well the observed V960 Mon spectrum of the early and the late phases (Figure \ref{fig:speccomp}). 
For the early phase, we found that the spectrum of HD 108968 in the archival data of UVES Paranal Observatory Project (\cite{2003Msngr.114...10B}) was nearly equal to our observed spectrum of V960 Mon. 
For the late phase, HD 185018 observed at Okayama Astrophysical Observatory/High Dispersion Echelle Spectrograph showed a good fit. 
Since absorption lines of these giants are narrow due to  their slow rotation, we smoothed their spectra to fit the spectrum of V960 Mon in the respectable phases. 

The fraction of the disk component in the observed spectrum of V960 Mon is able to be estimated from the brightness increment of V960 Mon.
The brightness increment can be estimated by comparing the brightness of the early and late phase to that of the pre-outburst phase. 
According to the past photometric observations reported in AAVSO, the approximate brightness increment in 6500~{\AA} is estimated as $\sim$ 2.8 mag and $\sim$1.8 mag in early and late phases, respectively, compared to the pre-outburst phase. 
Hence, the fraction of the disk component in the observed spectrum of V960 Mon was estimated as $\sim$ 92\% and $\sim$ 80\% in the early and late phases, respectively. 
Based on the estimated fraction, we subtracted the disk component in the spectrum of V960 Mon. 

The H$\alpha$ line in the spectrum of which the disk component was subtracted show a strong emission feature and blue shifted absorption feature. 
We found the \textit{EW} of the H$\alpha$ emission to be roughly $-$25~{\AA} in the disk component subtracted spectra, both in the early and the late phases. 
The \textit{EW} error was estimated as $\sim$3 {\AA}, assuming a 0.1 magnitude uncertainty in the brightness increment. 
This implies that the emission feature in H$\alpha$ was generally stable during the observation period.

\citet{2015ApJ...806L...4C} argued that V960 Mon has possibly two companions, which are young stellar objects, and described that Pa$\beta$ and Br$\gamma$ emissions observed in the near-infrared spectrum are emitted from these companions. 
Therefore, it is possible that the H$\alpha$ emission also originates from these companions. 
Since the increased brightness of V960 Mon was relatively small, there is a possibility that the emission feature of the companions are relatively easy to observe compared to other FUors in binary system.
On the other hand, a large fluctuation in the blueshifted absorption feature can also affect the \textit{EW} of the emission. 
As \citet{2012MNRAS.426.3315P} conducted in the case of FU Ori, further investigation with spectra of high spectroscopic resolution, high S/N, and high time resolution are needed to  give more constraints on the evolution of the disk and wind structures of V960 Mon.

\section{Summary} \label{sec:sum}

We conducted spectroscopic monitoring observations of the FU Ori type star V960 Mon for two years from 2015 January to 2017 January. 
The \textit{EW}s of the absorption and the emission lines were measured. 
The variations of the \textit{EW}s of the Fe~{\footnotesize I} and Fe~{\footnotesize II} lines show that $T_{\mathrm{eff}}$ decreased and that $g$ of the luminous source increased. 
This result suggests that the luminous source of the V960 Mon is the accretion disk and that the mass accretion rate decreased, the latter of which caused $T_{\mathrm{eff}}$ to decrease. 
The increase in $g$ reflects the evolution of the vertical structure in the accretion disk. Emission lines were also detected in our observations, which may be emitted from the nearby companions. 
Further observations with high spectroscopic resolution and high time resolution is needed to investigate the evolution of the disk and the wind structure with high accuracy.

\medskip

We used the data obtained at Okayama Astrophysical Observatory, which had been collected through SMOKA operated by the Astronomy Data Center, National Astronomical Observatory of Japan. 
We also used the data from the UVES Paranal Observatory Project (ESO DDT Program ID 266.D-5655). 
This work was supported by JSPS KAKENHI Grant Numbers 26103708, 26870507, and 25870124. 

\software{IRAF \citep{1986SPIE..627..733T,1993ASPC...52..173T}, SPTOOL (http://optik2.mtk.nao.ac.jp/\~{}takeda/sptool/)}. 

\bibliographystyle{aasjournal}
%\bibliography{reference}

\end{document}